\documentclass[preprint,aps,showpacs,pre,amsmath]{revtex4}
\usepackage{epsfig}

\begin{document}

\title{Touching points in the energy band structure of bilayer graphene superlattices}

\author{ C. Huy Pham$^{1,2}$ and V. Lien Nguyen$^{1,3}$
         \footnote{Corresponding author, E-mail:
         nvlien@iop.vast.ac.vn} }
\affiliation{ $^1$Theoretical and Computational Physics Department., Institute of Physics, VAST,  \\
10 Dao Tan, Ba Dinh Distr.,  Hanoi 10000,  Vietnam \\
$^2$ SISSA/International School for Advanced Study, Via Bonomea 265, I-34136 Trieste, Italy \\
$^3$ Institute for Bio-Medical Physics, 109A Pasteur, $1^{st}$ Distr., Hochiminh City, Vietnam }

\vspace*{1cm}
\begin{abstract}
Energy band structure of the bilayer graphene superlattices with zero-averaged periodic $\delta$-function potentials are studied within the four-band continuum model. Using the transfer matrix method, studies are mainly focused on examining the touching points between adjacent minibands. For the zero-energy touching points the dispersion relation derived shows the Dirac-like double-cone shape with the group velocity which is periodic in the potential strength $P$ with the period of $\pi$ and becomes anisotropic at relatively large $P$. From the finite-energy touching points we have identified those located at zero wave-number. It was shown that for these finite-energy touching points the dispersion is direction-dependent in the sense that it is linear or parabolic in the direction parallel or perpendicular to the superlattice direction, respectively. We have also calculated the density of states and the conductivity that demonstrate a manifestation of the touching points of interest.
\end{abstract}

\maketitle
\section{Introduction}
Bilayer graphene shares with monolayer graphene many properties holding a very high potential for electronics applications such as the excellent electric and thermal conductivities at room temperature or a possibility to control the electronic structure externally (see for review Refs.\cite{neto,castro,cann}). However, bilayer graphene (BLG) also exhibits the unique features that make it qualitatively distinct from monolayer graphene (MLG). For instance, the integer quantum Hall effect in BLG indicates the presence of massive chiral quasi-particles with a parabolic dispersion at low energies (rather than massless quasi-particles with a linear dispersion in MLG). From application point of view the ability to open a gap in the BLG energy spectrum and to turn it flexibly by an external electric field is exclusively important. It is expected that the BLG-based nano-devices would show the functionalities different from those in the corresponding MLG-based devices. Moreover, with only two layers in structure the BLG represents the thinnest possible limit of a large class of intercalated materials which are recently attracting much attention from condensed matter physicists as well as material science researchers \cite{geim}.

As is well-known, an external periodic potential can essentially modify the energy band structure of materials resulting in unusual transport properties. The energy band structure of MLG under a periodic potential (monolayer graphene superlattice - MLGSL) has been extensively studied in a number of works for the potentials of different natures (electrostatic \cite{park,brey,barbier,huy} or magnetic \cite{ghosh,masir,snyman,dellan,lequi}) and different shapes (Kronig-Penny \cite{park,huy,masir,lequi}, cosine \cite{brey} or square \cite{barbier}). Interesting discoveries have been reported, such as a strongly anisotropic renormalization of the carrier group velocity and an emergence of extra Dirac points (DPs) in the band structure of electrostatic MLGSLs \cite{park,brey,barbier,huy} or an emergence of finite-energy DPs in the band structure of magnetic ones \cite{snyman,dellan,lequi}. Much less work has been devoted to the BLG-superlattices (BLGSLs) \cite{peeters,killi,tan}. Barbier, Vasilopoulos, and Peeters (BVP) introduced the Kronig-Penny model of BLGSLs with $\delta$-function potentials and predicted that either a pair of zero-energy touching points is generated or a direct band gap is opened in the energy spectrum, depending on the strength of the $\delta$-function potentials \cite{peeters}.

In the present paper we study the energy band structure of BLGSLs within the same model of zero-averaged periodic $\delta$-function potentials as that stated by BVP \cite{peeters} (hereafter referred to as BVP-model for short). Our study is however focused on examining the touching points between adjacent minibands, including the zero-energy touching points claimed before in ref.\cite{peeters} and the finite-energy touching points identified first in this work.

The paper is organized as follows. In Sec.2 we briefly describe the problem  under study within the four-band continuum model and supplement the zero-energy touching points claimed before with the dispersion that shows the Dirac-like double-cone shape and the group velocity which is periodic in the potential strength $P$ with period of $\pi$ and becomes strongly anisotropic at relatively large $P$. In Sec.3 we show that for any finite potential strength there always exist the finite-energy touching points at zero wave number $(k_x = 0, k_y = 0)$, regardless of whether there are zero-energy touching points or whether there is a band gap. Impressively, the dispersion associated to these touching points is direction-dependent in the sense that in one direction the dispersion is linear whereas in other direction it is parabolic. Sec.4 presents the density of states and the conductivity that demonstrate possible manifestations of the touching points of interest. The paper is closed with a brief summary in Sec.5.
\section{Four band Hamiltonian and zero-energy Dirac points}
We consider BLGSLs arising from an infinitely flat Bernal-stacked BLG in a periodic one-dimensional potential $V(x, y) \equiv V(x)$. Within the continuum nearest-neighbor, tight-binding model the four-band Hamiltonian describing low-energy excitations near one Dirac point (say, $K$) for these BLGSLs has the form:
\begin{equation}
  H \ = \ \left(  \begin{array}{cccc}
        0  & \ v_F \hat{\pi}  &  \ t_\perp  &  \ 0  \\
        v_F \hat{\pi}^+  &  \ 0 &  \ 0  &  \ 0       \\
        t_\perp  &  \ 0  &  \  0  &  \ v_F \hat{\pi}^+  \\
        0  &  \ 0  &  \ v_F \hat{\pi}   &  \  0
    \end{array}   \right) \  + \ V(x) {\cal I} \ ,
\end{equation}
where  $\hat{\pi} = p_x + i p_y$, $p = (p_x , p_y )$ is the in-plane momentum, $v_F = \sqrt{3} t a / (2 \hbar ) \approx 10^6 \ m/s$ is the Fermi velocity, $t \approx 3 \ eV$ is the intralayer nearest-neighbor hopping energy, $a = 2.46 \AA$ is the lattice constant of graphene, $t_\perp \approx 0.39 \  eV$ is the interlayer nearest-neighbor hopping energy, and ${\cal I}$ is the identity matrix. This Hamiltonian is limited to the case of symmetric on-site energies. The potential $V(x)$ under study is the zero-averaged periodic $\delta$-function potential with strength $P$ and period $L$ defined as [see Fig.1$(a)$]:
\begin{equation}
V(x) \ = \ P \{ \ \sum_n \delta (x - n L) - \sum_n \delta [x - (n + 1/2) L] \ \}.
\end{equation}

In fact, the model studied is exactly the one introduced by BVP (BVP-model) \cite{peeters}. These authors claimed that in studying the energy band structure of (at least) BLGSLs with the $\delta$-function potentials the two-band approximation [see for example \cite{cann}] is not accurate enough and the four-band Hamiltonian of eq.(1) should be applied.

\begin{figure} [h]
\begin{center}
\includegraphics[width=14.0cm,height=9.0cm]{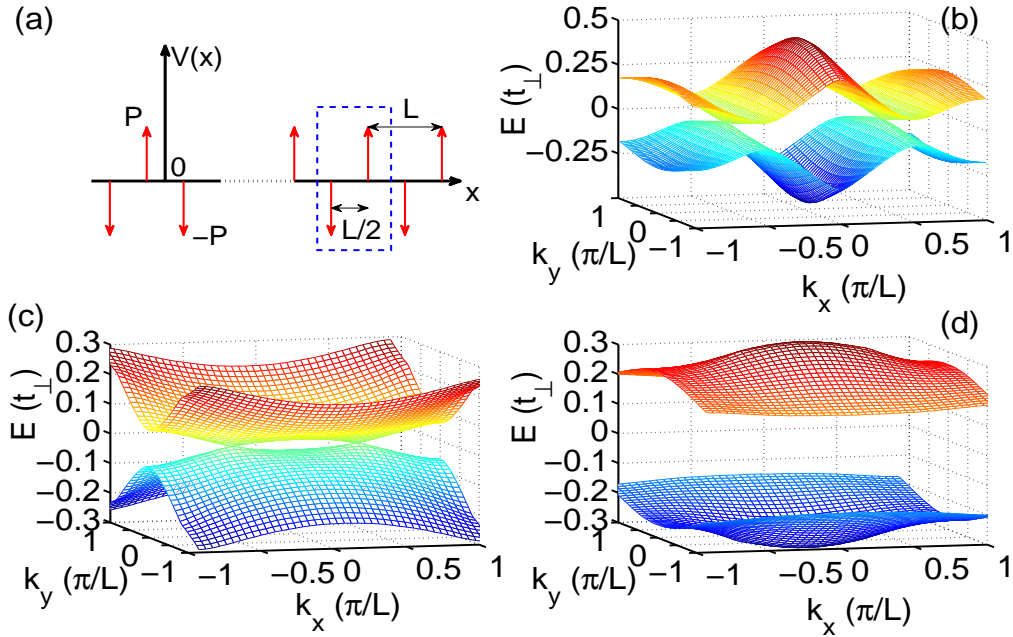}
\caption{
$(a)$ Schematics of a zero-averaged periodic $\delta$-function potential $V(x)$ with strength $P$ and period $L$ (The dashed-line box describes the unit cell in the T-matrix calculation); $(b)$, $(c)$, and $(d)$ are the energy spectra of the lowest conduction and highest valence minibands for BLGSLs with $[L = 3, P = 1.5]$, $[L = 8, P = 0.1 \pi ]$, and $[L = 8, P = 0.4 \pi ]$, respectively. For $L = 8 > L_C$ either a pair of zero-energy DPs is generated $(c)$ or a direct band gap opens up $(d)$, depending on $P$. For $L < L_C$ a pair of zero-energy DPs is always existed, regardless of $P$.}
\end{center}
\end{figure}

Due to a periodicity of the potential $V(x)$ [eq.(2)] the time-independent Schr\"{o}dinger equation $H \Psi = E \Psi$ for the Hamiltonian $H$ of eq.(1) could be most conveniently solved using the transfer matrix method \cite{peeters,chau}, which generally reduces the energy spectrum problem to solving the equation [see Appendix]:
\begin{equation}
det \ [ \ T \ - \ e^{i k_x L} R^{-1}_I (L) \ ] \ = \ 0 ,
\end{equation}
where $k_x$ is the Bloch wave vector (along the $x$-direction of the periodic potential with the period of $L$) and $T$ and $R_I$ are matrices, depending on the Hamiltonian of interest [see Appendix].

In the case of MLGSLs, when the Hamiltonian $H$ and, therefore, $T$ and $R_I$ are $2 \times 2$ matrices, equation (3) can be analytically solved that gives straightaway a general expression for the dispersion relation $E(\vec{k})$  \cite{huy,lequi}. For BLGSLs in the four-band model of eq.(1), the $T$-matrix calculations have been in detail described in Ref.\cite{peeters}, however the equation (3) with $(4 \times 4)$-matrices $T$ and $R_I$ becomes too complicated to be solved analytically and therefore, in general, the dispersion relation can not be derived explicitly. We have numerically solved eq.(3) and show in Fig.1$(b-d)$ the lowest conduction and the highest valence minibands obtained for some values of the potential parameters $P$ and $L$. Hereafter, we introduce the dimensionless variables: $E \rightarrow E / t_\perp$, $V \rightarrow V / t_\perp$, $x(L) \rightarrow x(L) / (\hbar v_F / t_\perp )$, and $k_{x(y)} \rightarrow k_{x(y)} / (t_\perp / \hbar v_F )$ with $t_\perp$ and $v_F$ given above.

The most impressive feature observed in Fig.1 is that, instead of the original zero-energy Dirac point (DP) of the pristine BLGs at $\vec{k} = 0$ there appeared a pair of new touching points in the $(k_y = 0)$-direction [Figs.1$(b)$ and $(c)$] or even a direct band gap [Fig.1$(d)$], depending on the values of $P$ and $L$. Actually, such a picture of the zero-energy touching points or the band gap has been reported before by BVP \cite{peeters} and it is here included for additional discussions. Before going over to detailed considerations we would like here to note that due to the symmetry of the periodic potential $V(x)$ of eq.(2) (with a zero spatial average) the energy spectrum of the BLGSLs under study should be symmetric with respect to the sign of energy that results in a double-cone dispersion in the vicinity of all the zero-energy touching points existed, and therefore these touching points could be referred to as the zero-energy DPs. We will return to this point later.

Next, in order to understand the new zero-energy DPs generated in the $(k_y = 0)$-direction (as seen in Figs.1$(b)$ and $(c)$) or the direct band gap opened at some values of $L$ and $P$ (as seen in Fig.1$(d)$), following BVP \cite{peeters} we consider the energy spectrum along the $(k_y = 0)$-direction. In this case, in the wave function $\Psi = ( \psi_{A_1},\psi_{B_1}, \psi_{B_2},\psi_{A_2} )^T$ the two components relating to the first layer $(\psi_{A_1},\psi_{B_1})$ and those to the second layer  $(\psi_{B_2},\psi_{A_2})$ become decoupled and therefore the energy spectrum can be obtained in the form of the transcendental equations \cite{note01}:
\begin{equation}
\cos (k_x L) \ = \ \cos^2 ( k_n L / 2) - \sin^2 ( k_n L / 2) D_n , \ \ n = 1 \ {\rm and} \ 2 ,
\end{equation}
where $D_n = \cos^2 P + [( E^4  + k_n^4 )/( 2 E^2 k_n^2 )] \sin^2 P$ and $k_n = \sqrt { E^2 - (-1)^n E }$. (In these equations all variables are dimensionless as defined above).
\begin{figure} [h]
\begin{center}
\includegraphics[width=12.0cm,height=9.0cm]{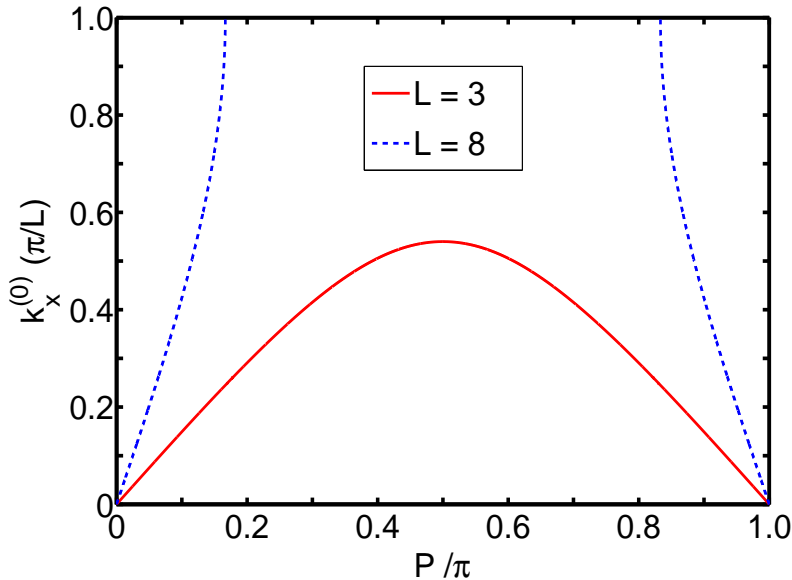}
\caption{
$k_x^{(0)}$ of eq.(5) [in unit of $\pi / L$] is plotted as a function of $P / \pi$ in two cases: $L = 3$ (red solid line) and $L = 8$ (blue dashed line).  $k_x^{(0)}$ is periodic in $P$ with the period of $\pi$.}
\end{center}
\end{figure}

We are interested in how the original zero-energy DP in the energy spectrum of the pristine BLG responds to the external periodical potential $V(x)$. To this end, because of the symmetry of the energy spectrum with respect to the zero-energy plane it is possible to identify the zero-energy DPs by taking the limit $E \rightarrow 0$ in the dispersion relation. Indeed, in this limit any of the equations (4) gives to possible zero-energy DPs  the $k_x$-coordinate that depends on the periodic potential parameters as
\begin{equation}
k_x \ = \ \pm \ k_x^{(0)}(P, L)  \ = \ \pm \ \arccos [1 - (L^2 / 8) \sin^2 P ] / L .
\end{equation}
This equation yields the real values for $k_x^{(0)}$ and therefore identifies the position of zero-energy DPs only if the strength $P$ and the period $L$ of the potential $V(x)$ obey the following condition:
\begin{equation}
| 1 - (L^2 / 8) \sin^2 P | \ \leq \ 1 \ \ {\rm or} \ \ L^2 \sin^2 P \ \leq \ 16
\end{equation}
Whenever this condition is fulfilled, instead of a single zero-energy DP at $\vec{k} = 0$ (i.e. the K-point) in the energy spectrum of the pristine BLG, the periodic potential $V(x)$ induces a pair of new zero-energy DPs located symmetrically at $k_x = \pm k_x^{(0)}$ along the $(k_y = 0)$-direction. A violation of the condition (6) means there exists no zero-energy DPs at all or, in other words, a direct band gap should be opened instead.

Note that the condition of eq.(6) is always fulfilled for $L < L_C = 4$. So, for any BLGSL with such a small potential period, $L < L_C ( \approx 6.75 \ nm \ {\rm given} \ t_\perp \approx 0.39 \ eV)$, there always has in the energy spectrum a pair of zero-energy DPs located at $(k_x =  \pm \ k_x^{(0)}, k_y = 0)$, regardless of the potential strength $P$. This is the case shown in Fig.1$(b)$ for the BLGSL with $L = 3$ and $P = 1.5$. Changing $P$ in this figure does not remove the pair of zero-energy DPs, but only shift their $k_x$-coordinates.

On the contrary, for any BLGSL of $L > L_C$, with increasing $ P $, the two zero-energy DPs, generated initially near the $(\vec{k} = 0)$-point, move away from this point in opposite directions along the $(k_y = 0)$-direction [Fig.1$(c)$]. At the potential strength $P = P_C$ defined by the upper limit in the condition of eq.(6), $\sin^2 P_C = 16 / L^2$, these DPs reach the edge of the Brillouin zone at $k_x = \pm \pi / L$. Further increase in $P$ opens a direct band gap [Fig.1$(d)$].

In order to see the whole evolution picture of the zero-energy DP positions and/or band gap as $P$ varies, we remark the two properties of the $k_x^{(0)}(P)$-function defined in eq.(4): $(i)$ $k_x^{(0)}(P) = k_x^{(0)} (P \pm n \pi )$ with $n$ integer number and $(ii)$ $k_x^{(0)}(P) = k_x^{(0)}( \pi - P)$. The former property simply means that the whole $k_x^{(0)}(P)$-picture is periodic in $P$ with the period of $\pi$ and therefore it is enough to examine $k_x^{(0)}$ in one period, $P \in [0, \pi]$. The latter shows the symmetry of the $k_x^{(0)}(P)$-picture in one period with respect to the middle point $P = \pi / 2$. These symmetries can be clearly seen in Fig.2 where $k_x^{(0)}$ of eq.(5) is plotted versus $P$ in a period for $L = 3$ (red solid line) and $L = 8$ (blue dashed line). In the case of $L = 3 < L_C$ the solid line shows a continuous and regular oscillation of $k_x^{(0)}$ between the minimum of zero at $P = n \pi$ and the maximum of  $ \pi / 2 L$ at $P = \pi / 2$. In the case of $L = 8 > L_C$ the dashed line shows the $P$-dependence of $k_x^{(0)}$ in the two symmetrical regions, $0 < P < P_C$ [$\approx 0.167 \pi $ in Fig.2] and $\pi - P_C < P < \pi$, when the zero-energy DPs are surviving. For the potential strengths in the middle region, $P_C < P < \pi - P_C$, a direct band gap opens up.

Note that the picture similar to the dashed line in Fig.2 was previously presented together with the band gap size in Fig.12 in ref.\cite{peeters}, where, however, the only  case of $L = 10 \ nm > L_C$ is discussed. Note also that the relations of eqs.(5) and (6) as well as the results presented in Figs.1 and 2 are exactly the same as those reported in ref.\cite{peeters}, though, as mentioned above \cite{note01}, the equations (3) and the corresponding equations (31) in the reference cited are unidentified yet.

\begin{figure} [h]
\begin{center}
\includegraphics[width=12.0cm,height=9.0cm]{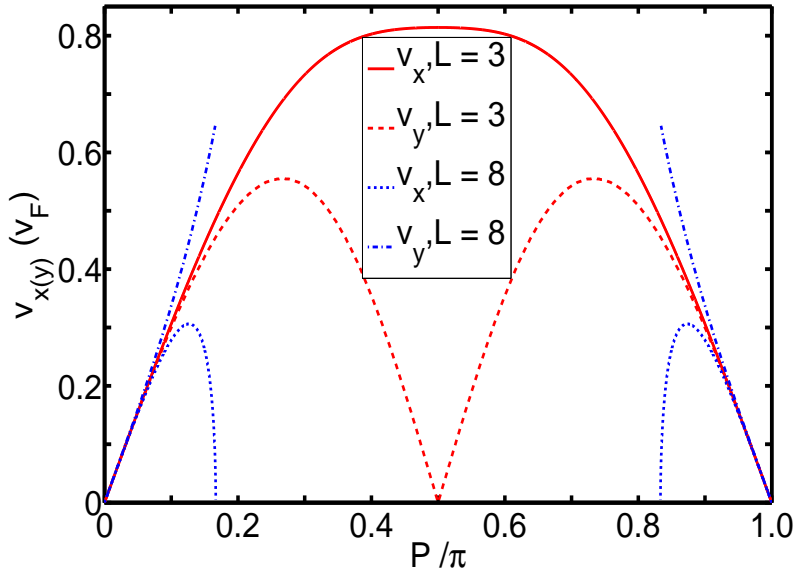}
\caption{
For zero-energy DPs: velocities $v_x$ of eq.(8) and $v_y$ (numerically calculated) [in units of $v_F$]  as the functions of the $P / \pi$ for BLGSLs with: $L = 3$ [$v_x$ - red solid line, $v_y$ - red dashed line] and $L = 8$ [$v_x$ - blue dotted line, $v_y$ - blue dash-dotted line]. Velocities are periodic in $P$ with the period of $\pi$.
}
\end{center}
\end{figure}

Further, to search for more accurate understandings of the new zero-energy DPs discussed we expand eq.(3) to the lowest order in $E$ and $k_x$ in the vicinity of these DPs that reveals the dispersion relation
\begin{equation}
E^2 \ = \ (k_x - k_x^{(0)})^2 v_x^2 \ + \ k_y^2 v_y^2 ,
\end{equation}
where the velocity $v_x$ can be deduced by expanding eq.(4):
\begin{equation}
v_x =  \frac{ \sin P \sqrt{ 16 - L^2 \sin^2 P }}{ 4 - (L^2 / 12) \sin^2 P },
\end{equation}
whereas $v_y$ could be numerically calculated. The Dirac-like form of the dispersion relation of eq.(7) explains why the zero-energy touching points could be referred to as the DPs.

Fig.3 presents $v_x$ and $v_y$ in one period of $P$ for the BLGSLs with periods of $L = 3 < L_C$ ($v_x$ - red solid and $v_y$ - red dashed line) and $L = 8 > L_C$ ($v_x$ - blue dotted and $v_y$ - blue dash-dotted line). Noticing that like $k_x^{(0)}$ these velocities are periodic in $P$ with the period of $\pi$ and, additionally, in one period they are symmetric with respect to the middle point $P = \pi /2$, it is enough to analyze the picture in a half of the period. Clearly, in both the cases under consideration the two curves $v_x (P)$ and $v_y (P)$ are practically coincided at small $P$, indicating an isotropy of the Dirac cones in this region of $P$-values. However, with increasing $P$, two velocities become largely separated, showing a strong anisotropy of the Dirac cones in the cases of  large potential strength $P$. Thus, Fig.3 demonstrates an interesting feature of the new zero-energy DPs: given a potential period $L$ the dispersion cone is practically isotropic if the potential strength $P$ is small and becomes strongly anisotropic at large $P$ .(Actually, the value $P_s$ below which the dispersion could be considered isotropic depends slightly on $L$:  in Fig.3 $P_s \approx 0.08 \pi$ or  $ 0.06 \pi$ for $L = 3$ or $L = 8$, respectively).

Fig.3 also demonstrates that for small $P$ [$P \leq 0.06 \pi$] the velocities are not only isotropic, but also almost independent of $L$. At larger $P$, there is a strong difference in the velocity behavior between the BLGSLs with $L > L_C$ and those with $L < L_C$. For the former BLGSLs, $v_y$ increases straight, while $v_x$ goes back to zero with the opening of the band gap. For the latter, when the gap is totally absent, $v_x$ reaches the maximum, while $v_y$ goes to zero at $P = \pi / 2$, implying that the dispersion turns to be one-dimensional at this value of $P$. Such a difference in the velocity behavior should find itself reflected in the transport properties.

It should be now mentioned that as a consequence of the pseudospin structure of the wave function in BLGs, a pair of new zero-energy DPs appeared at small potential strengths or a direct band gap opened at higher ones have been also reported in the energy spectra of BLGSLs with different periodic potential shapes, rectangular potential \cite{killi} or sine potential \cite{tan}, and perhaps could be seen as the common features in the energy band structure of all BLGSLs with zero-averaged periodic potentials. The detailed behavior of the zero-energy DPs and/or the band gap is however depending on the potential shape. In particular, for the sine potential within the two-band continuum model Tan {\sl et al.} demonstrated an emergence of new zero-energy DPs even along the $(k_x = 0)$-direction at higher values of the potential strength [see Fig.3 in Ref.\cite{tan}]. It is worthy to note that such the zero-energy DPs at $k_x = 0$ should not exist in the band structure of the BLGSLs with periodic $\delta$-function potentials we are interested in \cite{note02}.
\section{Finite-energy touching points}
In Fig.4 we plot the cuts of the band structure along the $(k_y = 0)$-plane, calculated numerically from eq.(3) for various values of the potential strength $P$ [ \ $0.1 \pi$ in $(a)$ and $(c)$; $0.4 \pi$ in $(b)$ and $(d)$ \ ] or the potential period $L$ [ \ 3 in $(a)$ and $(b)$; 8 in $(c)$ and $(d)$ \ ]. This figure is focused on showing several minibands next to the lowest one. (Due to the symmetry of energy spectra with respect to the zero-energy plane only the positive energies are shown). Interestingly, in all boxes with different $P$ and/or $L$ in Fig.4 apart from the zero-energy DPs described in the previous section there appeared to see the degeneracy points at finite energies, the touching points of adjacent minibands. Such the touching points do exist even in the cases when there is no zero-energy DP at all in the energy spectrum as can be seen in the box $(d)$ for $L= 8$ and $P = 0.4 \pi$.

It seems that among the finite-energy touching points observed in Fig.4 there is a class of touching points which could be exactly identified. Indeed, equations (4) always have solutions with energies, corresponding to the equality $\sin (k_n L/2) = 0$. These energies are then determined as
\begin{equation}
E_\nu^{(i,j)} \ = \ \frac{1}{2} \ [ \  (-1)^i  + (-1)^j  \sqrt{ 1 + 16 \nu^2 \pi^2 / L^2 } \ ], \ i, j = 1, 2;  \ \ \nu = 1, 2, 3, \cdots
\end{equation}
which are the energy positions of all possible finite-energy touching points located at $(k_x = 0, k_y = 0)$ [some of these points are marked by the red square in Fig.4]. Note that the energies $E_\nu^{(i,j)}$ (9) only depends on $L$, so in the energy spectra of all BLGSLs of a given $L$ the energy-positions of the touching points, associated with the same indexes $\nu , i$ and $j$, are exactly coincident. (Compare the marked touching points in the two boxes in the same line in Fig.4).

Generally, to examine the dispersion behavior at a touching point we should expand eq.(3) in the vicinity of this point. As a demonstration, we consider the lowest from all the touching points of interest, the point $E_1^{(1,2)} = [ - 1 + \sqrt{1 + 16 \pi^2 / L^2 } ] / 2 $ (the lowest touching point marked by a square  in all the boxes in Fig.4). Keeping $k_y = 0$, we expand eq.(4) in the vicinity of the point $( E = E_1^{(1,2)} , k_x = 0 )$ that  gives
\begin{equation}
E \ - \ E_1^{(1,2)} \ = \ \pm \ k_x v_x ,
\end{equation}
where
\begin{equation}
v_x \ = \ [ ( L^2 / 16 \pi^2 + 1 ) ( L^2 \sin^2 P / 16 \pi^2 + 1 ) ]^{1/2} .
\end{equation}
On the other hand, keeping $k_x = 0$ we can expand eq.(3) in the vicinity of the point $(E = E_1^{(1,2)}, k_y = 0)$. This can be done only numerically, writing eq.(3) in the form of $f(E, k_y ) = 0$ [see for detail Refs. \cite{huy,lequi}]. It seems that the function $f$ has the zero first derivative, $\partial f / \partial k_y = 0$, so it can be readily written:
\begin{equation}
E \ - \ E_1^{(1,2)} \ = \ k_y^2 / 2 m_\pm \ ,
\end{equation}
where $m_\pm$ being the masses of the parabolic dispersions [the signs $\pm$ at $m$ are corresponding to  $\pm$ in eq.(10)].
\begin{figure} [h]
\begin{center}
\includegraphics[width=12.0cm,height=9.0cm]{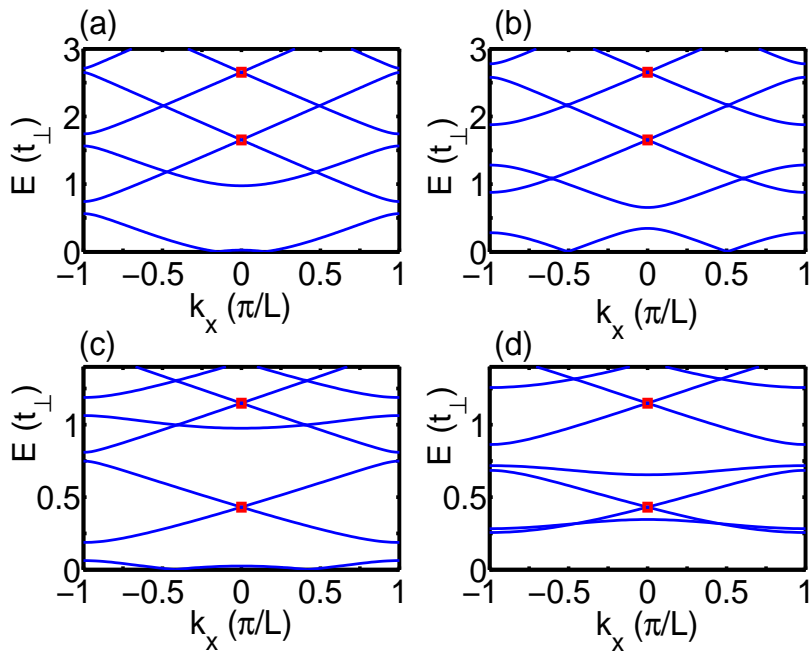}
\caption{
Cuts of the band structure along the $(k_y = 0)$-plane for BLGSLs with different $L$ and $P$: $(a)$, $(b)$, $(c)$, and $(d)$ for $[L, P] = [3, 0.1 \pi ], \ [3, 0.4 \pi ], \ [8, 0.1 \pi ]$, and $[8, 0.4 \pi ]$, respectively. The lowest finite-energy touching points of eq.(9) are marked by the red squares (The spectrum is symmetric with respect to the sign of the energy and only the positive energies are shown). The energy $E$ and the wave number $k_x$ are in units of $t_\perp$ and $\pi / L$, respectively. Note, except the finite-energy touching points marked there are also those at $k_x \neq 0$.
}
\end{center}
\end{figure}

Thus, around the touching point located at $E = E_1^{(1,2)}$ the dispersion seems to be direction-dependent in the sense that it is (Dirac-like) linear in the $k_x$-direction [eq.(10)], but parabolic in the $k_y$-one [eq.(12)]. While the velocity $v_x$ in eq.(10) is well defined [eq.(11)], the manner in which the two minibands touch each other at $E_1^{(1,2)}$ is still associated with the masses $m_\pm$ in eq.(12). Unfortunately, we are unable to analytically estimate $m_\pm$, so we present in Fig.5 the numerical results of these masses plotted as a function of $P$ in two cases: $(a)$ $L = 3$ and $(b)$ $L = 8$. Surprisingly, Fig.5 shows $(i)$ the masses $m_\pm$ exhibit the same symmetric properties as $k_x^{(0)}$ in Fig.2 or $v_{x(y)}$ in Fig.3, $(ii)$ both $m_\pm$ may be either positive or negative, depending on $L$ and $P$ [Fig.5$(b)$], and $(iii)$ the two masses $m_+$ and $m_-$ change in very different ways as $P$ varies. In the case of $L = 3 < L_C$, when the band gap never appears, the two masses are both positive, but different in value (except the single point $P = 0.5 \pi$). In the case of $L = 8 > L_C$, when the band gap appears at $P_C < P <  \pi  - P_C$, both masses $m_\pm$ can change value and sign as $P$ varies. With increasing $P$ the two masses are all positive, but with different values at $P < P_C$. They become different in sign in some region of $P > P_C$ and then tend to be coincident at $P = \pi / 2$.
\begin{figure} [h]
\begin{center}
\includegraphics[width=12.0cm,height=9.0cm]{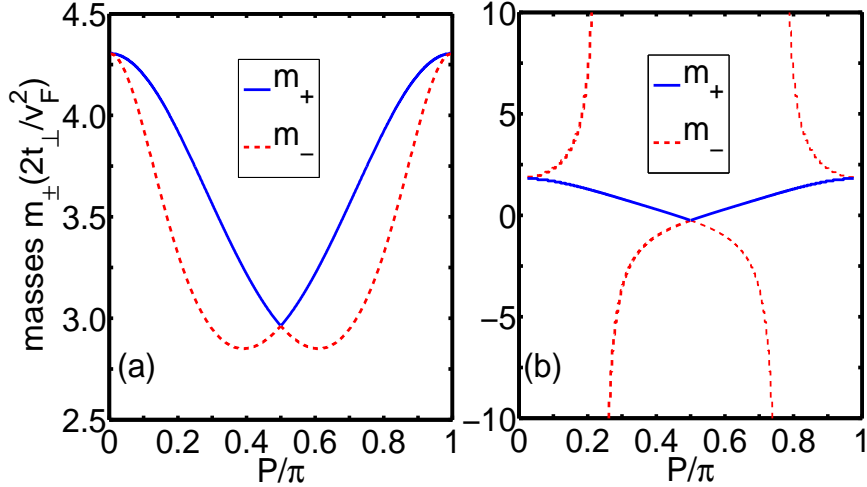}
\caption{
The masses $m_\pm$ in eq.(12) [in units of $( 2t_\perp / v_F^2  )$] for the touching point at $E = E_1^{(1,2)}$ periodically depend on $P$ with the period of $\pi$ in two cases: $L = 3$ $(a)$ and $L = 8$ $(b)$. In both cases (boxes): blue solid line describes $m_+$ and red dashed line - $m_-$.
}
\end{center}
\end{figure}

Due to the fact that the two masses $m_+$ and $m_-$ in the dispersion (12) associated with the touching point $E_1^{(1,2)}$ behave in very different ways towards the potential strength $P$, in general, this touching point should not be referred to as the DP. Actually, the dispersion relations of eqs.(10) and (12) are qualitatively applied for all the touching points located at $(k_x = 0, k_y = 0)$ and $E$ defined in eq.(9) [but with different $v_x$ and $m_\pm$ certainly] and,   therefore, all the properties belong to the lowest touching point at $E_1^{(1,2)}$ could be qualitatively considered general for the whole class of finite-energy touching points of interest.

In fact, the Dirac points with direction-dependent cones have been reported for various graphynes \cite{malko,soodch}. In particular, for the DP at the $M$-point in the spectrum of $\gamma$-graphyne the cone was shown to be direction-dependent in the way similar to eqs.(10) and (12): the dispersion is linear in the ($M - \Gamma $)-direction, but parabolic in the ($M - K$)-one [Fig.2 in Ref.\cite{soodch}].
\section{Density of states and conductivity}
\begin{figure} [h]
\begin{center}
\includegraphics[width=12.0cm,height=8.0cm]{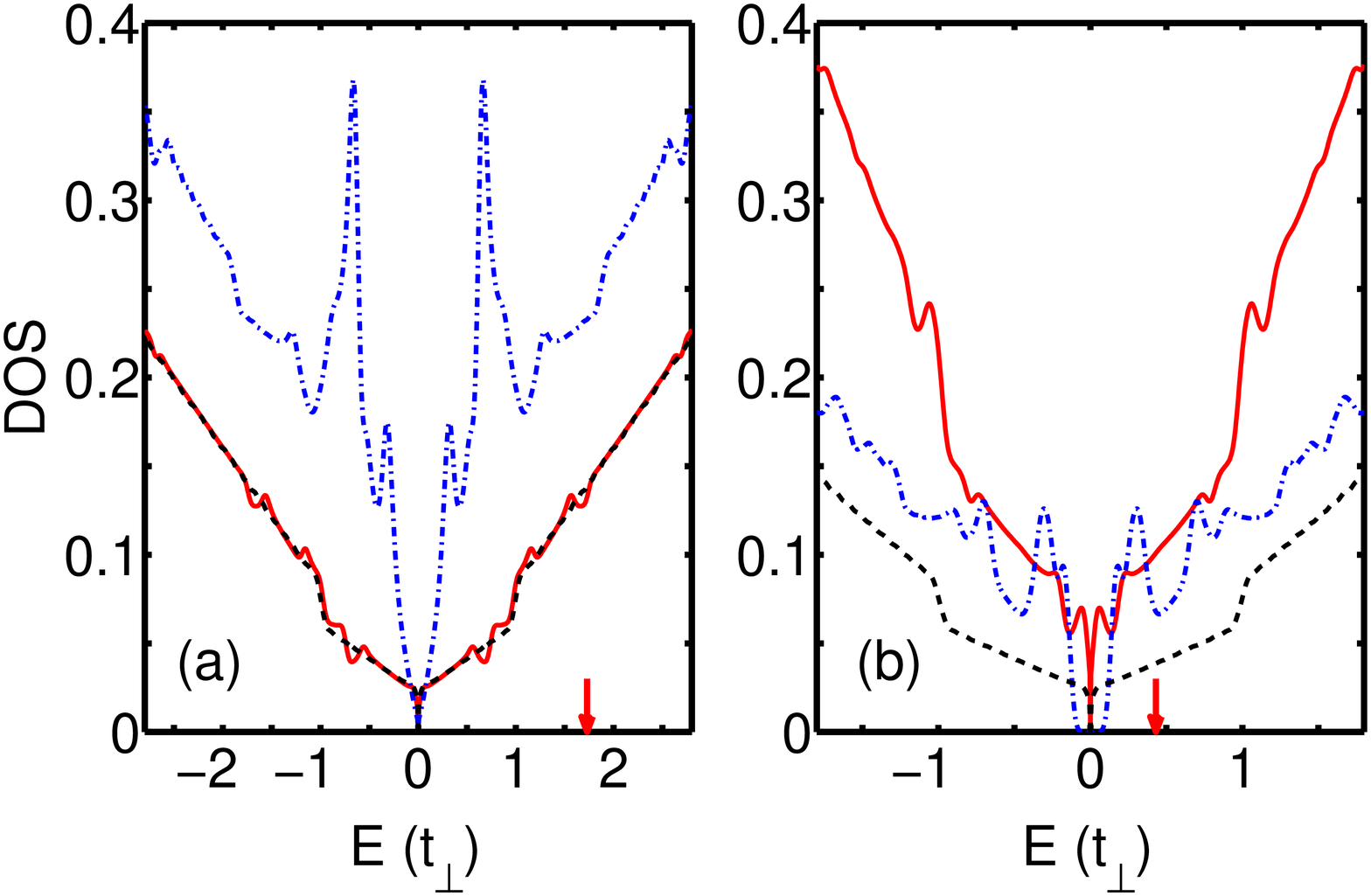}
\caption{
Density of states are shown for the BLGSLs with different $L$ and $P$:  $(a)$ $L = 3$ and $(b)$ $L = 8$; red solid lines - $P = 0.1 \pi$ and blue dash-dotted lines - $P = 0.4 \pi$. In both boxes the DOS for the pristine BLG (without periodic potential) is also shown for comparison (dashed line). Arrows indicate the energy  $E_1^{(1,2)}$ of the lowest finite-energy touching point of interest.
}
\end{center}
\end{figure}
With the energy band structures determined we calculated the density of states (DOS) and further the low temperature conductivity. Calculations have been performed in the same way as that suggested for MLGSLs in ref.\cite{masir}.

Fig.6 presents the DOS of the BLGSLs in two typical cases: $(a)$ $L = 3 < L_C$, when a pair of the zero-energy DPs is always existed, regardless of $P$ and $(b)$ $L = 8 > L_C$, when there exists either a pair of zero-energy DPs or a direct band gap, depending on $P$. In each box the three DOS-curves are shown for comparison: the dashed line for the pristine BLG ($P = 0$); the red solid line for $P = 0.1 \pi$; and the blue dash-dotted line for $P = 0.4  \pi$. The arrows indicate the energy-positions $E = E_1^{(1,2)}$ of the lowest finite-energy touching points defined in eq.(9). (This energy does not depend on $P$, so it is the same for both the red solid and the blue dash-dotted curves in each box). Notice that the DOSs are symmetric with respect to the zero-energy.
\begin{figure} [h]
\begin{center}
\includegraphics[width=12.0cm,height=9.0cm]{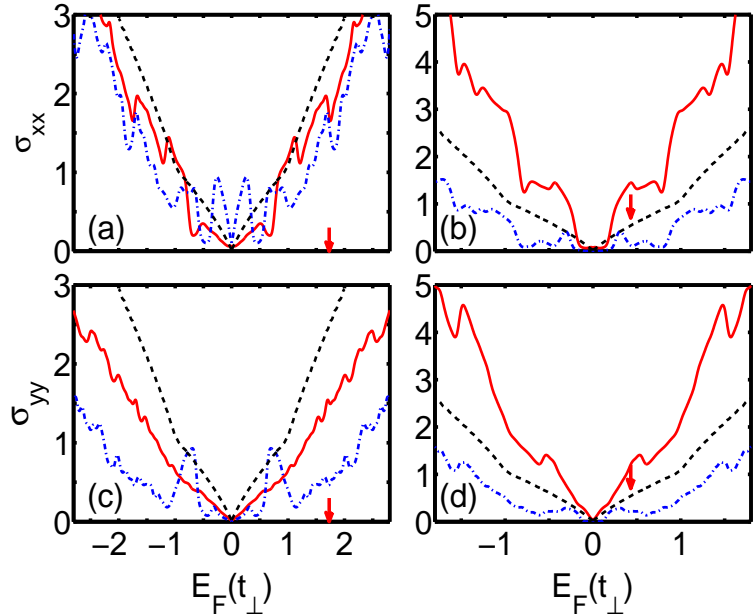}
\caption{
Conductivities $\sigma_{xx}$ [$(a)$ and $(b)$] and $\sigma_{yy}$ [$(c)$ and $(d)$] against the Fermi energy $E_F$ for the same structures with the DOS presented in Fig.6: $L = 3$ [$(a)$ and $(c)$] and $L = 8$ [$(b)$ and $(d)$]. In each box the curves present the conductivity  $\sigma_{xx}$ (or $\sigma_{yy}$) for BLGSLs with $P = 0.1 \pi$ (red solid line) or $P = 0.4 \pi$ (blue dash-dotted line), and for the pristine BLG (dashed line). Arrows indicate the energy $E_1^{(1,2)}$.
}
\end{center}
\end{figure}

In general, the DOSs of BLGSLs in Fig.6 display a fluctuating behavior, comparing to the DOS of the pristine BLG [dashed curves]. Notice that a similar fluctuation has been reported by BVP in the DOS of MLGSLs \cite{barbier}. The central minimum (at $E = 0$) is related to the zero-energy DPs or the band gap [the blue dash-dotted curve for $P = 0.4 \ \pi$ in $(b)$]. The local dips at finite energies are corresponding to the finite-energy touching points at $k_y = 0$ (either those at $k_x = 0$ identified above or those at $k_x \neq 0$ as can be seen  in Fig.4), whereas the peaks are located at the bending points between these touching points. In the left box for $L = 3$ the DOS-curve for $P = 0.1 \pi$ [red solid] though being very close to that for the pristine BLG exhibits a clear dip at the energy $E = E_1^{(1,2)}$. An increase of $P$ makes the DOS-curve more fluctuated, keeping a pair of DPs at $E = 0$ [see the blue dash-dotted curve for $P = 0.4 \pi$]. In the right box for $L = 8$, the DOS-curve for $P = 0.1 \pi$ [red solid] still demonstrates an existence of the zero-energy DPs, while that for $P = 0.4 \pi$ [blue dash-dotted] shows clearly a direct band gap [of $\approx 0.3 t_\perp$ in width] and a deep local minimum at $E = E_1^{(1,2)}$.

An accurate reflection of the DOS discussed in Fig.6 could be found in the conductivity. Fig.7 shows the conductivities $\sigma_{xx}$ [$(a)$ and $(b)$] and $\sigma_{yy}$ [$(c)$ and $(d)$] plotted versus the Fermi energy $E_F$ for the same BLGSLs as those analyzed in Fig.6, including the conductivities of the pristine BLG [the dashed curves, they are identical in all boxes]. The common features of the conductivities of all the BLGSLs studied in Fig.7 are $(i)$ the conductivities are symmetrical with respect to the sign of the Fermi energy. (Such a conductivity symmetry is a direct consequence of the symmetry of the DOS with respect to the sign of the energy.) and $(ii)$ both $\sigma_{xx}$ and $\sigma_{yy}$ strongly fluctuate against $E_F$ and, additionally, for a given $L$ the fluctuation in $\sigma_{xx}$ is stronger than that in $\sigma_{yy}$. (Such a conductivity fluctuation is resulted from the fluctuation in the DOS: the conductivity goes up (down) as the Fermi energy moves through a peak (dip) in the DOS).

Comparing the boxes in Fig.7 to each other reveals additional features as follows $(i)$ for BLGSLs with $L = 3 < L_C$  [$(a)$ and $(c)$] both the conductivities $\sigma_{xx}$ and $\sigma_{yy}$ are in average smaller than the conductivity of the pristine BLG and slightly decrease with increasing $P$; $(ii)$ for BLGSLs with $L = 8 > L_C$ [$(b)$ and $(d)$] the conductivities are much more (and unsystematically) sensitive to a change in $P$, showing the existence of a direct band gap in the case of $P = 0.4 \pi$; and $(iii)$ while $\sigma_{xx}$ for the BLGSLs with $L = 3$ [$(a)$] shows an impressive dip at the Fermi energy  corresponding to the lowest finite-energy touching point     $E_1^{(1,2)}$ (indicated by the arrow), for the BLGSLs with $L = 8$ the effect of this point in the conductivities is rather weaker in the case of $P = 0.4 \pi$ and almost invisible in the case of small $P$ [red solid curve for $P = 0.1 \pi$ in $(b)$].

We would like to recall that except the finite-energy touching points identified at $(k_x = 0, k_y = 0)$ there are also the finite-energy touching points located at $k_x \neq 0$ which should certainly affect the DOS and the conductivity behaviors.
\section{Conclusions}
We have studied  the energy band structure of the BLGSLs with zero-averaged periodic $\delta$-function potentials (arranged along the $x$-direction) within the framework of the four-band continuum model, using the transfer matrix method. In the band structures obtained an analysis has been focused on the touching points between adjacent minibands which produce a certain effect on the transport properties. For the zero-energy touching points at $k_y = 0$ claimed before we were able to derive the dispersion relation. The Dirac-like double-cone shape of the dispersion obtained provides that these zero-energy touching points could be really referred to as the Dirac points. On the other side, the group velocity shows that  the dispersion may be either isotropic or strongly anisotropic, depending on the potential strength. We have also noted that the direct band gap can be opened in the energy spectrum of only the BLGSLs with large enough potential periods ($L > L_C$).

From the finite-energy touching points we are able to exactly identify those located at $(k_x = 0, k_y = 0 )$. It was shown that in the vicinity of the finite-energy touching points identified the dispersion is direction-dependent in the sense that it is linear or parabolic in the $k_x$- or  $k_y$-direction, respectively. Additionally, numerical calculations show that the value as well as the sign of the "electron"- and the "hole"-masses in the parabolic dispersion in the $k_y$-direction may be very differently varied, depending on the periodic potential parameters. The touching points at zero- as well as finite-energies may find themselves reflected in the oscillating behavior of the density of states and then the conductivities which have been calculated for BLGSLs of different potential parameters. \\ \\
{\bf Acknowledgments} \\
This work was financially supported by Vietnam National Foundation for
Science and Technology Development under Grant No. 103.02-2013.17. We thank Ms. Nguyen Thi Thuong for helpful discussions. \\ \\
{\large \bf Appendix} \\ \\
We shortly describe how the central eq.(3) can be derived. Following the general idea of the $T$-matrix method \cite{chau}, we first consider the wave functions of the equation $H \psi = E \Psi$ in the regions of constant potential, $V(x) = V_0 = constant$. For $H$ of eq.(1) these functions can be generally written in the form $\Psi = Q R(x) [ A, B, C, D ]^T \exp (ik_y y )$. They can be then simplified by the linear transformation $Q \rightarrow {\cal T} Q$ with \cite{peeters}
\begin{equation}
{\cal T} \ = \ \frac{1}{2} \left(  \begin{array}{cccc}
                         1  \ & \ \ 0 \ & \ -1 \ & \ 0    \\
                         0  \ & \ \ 1 \ & \ 0  \ & \ - 1  \\
                         1  \ &  \ \ 0 \  & \  1 \ & \  0  \\
                         0  \ & \ \ 1 \  & \  0 \  & \ 1
            \end{array}    \right)   \nonumber
\end{equation}
So
\begin{equation}
\ \ \ \ \ \ \ \ \ \ \ \ \  Q \rightarrow {\cal T} Q \ = \ \left(  \begin{array}{cccc}
       1   \  & \  1 \ & \ 0 \ & \  0  \\
     k_1 /E' \ & \ -k_1 /E' \ & \ - i k_y /E' \ & \ - i k_y /E'  \\
       0 \ & \ 0 \ & \ 1 \ & \ 1   \\
     - i k_y /E' \ & \ - i k_y /E' \ & \  k_2 /E' \ & \ -k_2 /E'
              \end{array}   \right) ,  \ \ \ \ \ \ \ \ \ \ \ \   (A.1)  \nonumber
\end{equation}
while the ${\cal T}$-transformation does not change the matrix $R(x)$:
\begin{equation}
\ \ \ \ \ \ \ \ \ \ \ \ \ \ \ \ \ \ \ \ R(x) \ = \ diag \ [ \ e^{ik_1 x},  e^{-ik_1 x}, e^{ik_2 x}, e^{-ik_2 x} \ ] , \ \ \ \ \ \ \ \ \ \ \ \ \ \ \ \ \ \ \ \ \ \ \ \ \ \ \ \ \ \ \ \ (A.2)  \nonumber
\end{equation}
where $E' = E - V_0$ and $k_n = \sqrt{ E'^2 - (-1)^n E' }$ with $ n = 1, 2$.

Further, the amplitudes ${\cal A}_I$ of the wave function before an unit cell and those after it, ${\cal A}_F$, could be related to each other by the $T$-matrix
\begin{equation}
\ \ \ \ \ \ \ \ \ \ \ \ \ \ \ \ \ \ \ \ \ \ \ \ \ \ \ \ \ \ \ \ \ \ \ {\cal A}_F \ =  \ T (F, I ) {\cal A}_I . \ \ \ \ \ \ \ \ \ \ \ \ \ \ \ \ \ \ \ \ \ \ \ \ \ \ \ \ \ \ \ \ \ \ \ \ \ \ \ \ \ \ \ \ \ \ \ (A.3) \nonumber
\end{equation}
On the other hand, the Bloch's theorem states
\begin{equation}
\ \ \ \ \ \ \ \ \ \ \ \ \ \ \ \ \ \ \ \ \ Q_I R_I (x) {\cal A}_F \ = \ \exp (i k_x L ) Q_I R_I (x - L) {\cal A}_I , \ \ \ \ \ \ \ \ \ \ \ \ \ \ \ \ \ \ \ \ \ \ \ \ \ \ \ \ (A.4) \nonumber
\end{equation}
where $k_x$ is the Bloch wave number and $L$ is the period of $V(x)$.

Comparing eq.(A.3) and eq.(A.4) gives rise to the equation (3),
\begin{equation}
\ \ \ \ \ \ \ \ \ \ \ \ \ \ \ \ \ \ \ \ \ \ \ \ \ \ \ \ \ \ det \ [ \ T  -  e^{i k_x L} R^{-1}_I (L)  \ ]  \ = \ 0 . \ \ \ \ \ \ \ \ \ \ \ \ \ \ \ \ \ \ \ \ \ \ \ \ \ \ \ \ \ \ \ \ \ \ \ \ \ (A.5) \nonumber
\end{equation}
For the $\delta$-function potential $V(x)$ of eq.(2) the unit cell is described in Fig.1$(a)$ and the $T(F, I)$-matrix in eq.(A.5) is defined as
$$ T(F, I) \ = \ [R_I (L/2)]^{-1} Q_I^{-1} S(-P) S' S(P) Q_I  \ ,$$
where $Q_I$ and $R_I$ are respectively defined in eqs.(A.1) and (A.2) for $V_0 = 0$,  $S' = Q_I R_I (L/2) Q_I^{-1}$, and
$$ S(P) \ = \ \left(  \begin{array}{cccc}
                       \cos (P) \ & \ -i \sin (P) \ & \ 0 \ & \ 0  \\
                       -i \sin (P) \ & \ \cos (P) \ & \ 0 \ & \ 0  \\
                             0 \ & \ 0 \ & \ \cos (P) \ & \ -i \sin (P)  \\
                             0 \ & \ 0 \ & \ -i \sin (P) \ & \ \cos (P)
                       \end{array}  \right) \ . $$


\begin{thebibliography}{99}
\bibitem{note01} Notice that eqs.(4) in this work and the corresponding eqs.(31) in Ref. \cite{peeters} are unidentified, though all results deduced from them, including eqs.(5) and (6), are respectively coincided. Surprisingly, we failed to reproduce these results starting from eqs.(31) in Ref.\cite{peeters}.
\bibitem{note02} We have checked this problem for BLGSLs with rectangular potentials studied in ref.\cite{tan} and learned that for this potential there exist the zero-energy DPs in the $(k_x = 0)$-line at large potential magnitudes similar to those observed for the sine potential in ref.\cite{killi}
\end{thebibliography}
\end{document}